\documentstyle[epsfig,prl,aps]{revtex}
\begin{document}
\draft \twocolumn[\hsize\textwidth\columnwidth\hsize\csname
@twocolumnfalse\endcsname

\title{Regular electrically charged vacuum structures
 with de Sitter center\\
  in Nonlinear Electrodynamics  coupled to General Relativity}

\author{Irina Dymnikova}

\address{Department of Mathematics and Computer Science,
         University of Warmia and Mazury,\\
         \.Zo{\l}nierska 14, 10-561 Olsztyn, Poland; e-mail:
         irina@matman.uwm.edu.pl}

\maketitle

\begin{abstract}

We address the question of existence of regular spherically symmetric electrically
charged solutions in Nonlinear Electrodynamics coupled to General Relativity.
Stress-energy tensor of the electromagnetic field has the algebraic structure
$T_0^0=T_1^1$. In this case the Weak Energy Condition leads to the de Sitter
asymptotic at approaching a regular center. In de Sitter center of an electrically
charged NED structure, electric field, geometry and stress-energy tensor are regular
without Maxwell limit which is replaced by de Sitter limit: energy density of a field
is maximal and gives an effective cut-off on self-energy density, produced by NED
coupled to gravity and related to cosmological constant $\Lambda$.
 Regular electric solutions satisfying WEC, suffer from one cusp in
the Lagrangian ${\cal L}(F)$, which creates the problem in an effective geometry whose
geodesics are world lines of NED photons. We investigate propagation of photons and
show that their world lines never terminate which suggests absence of singularities in
the effective geometry.To illustrate these results we present the particular example
of the new exact analytic spherically symmetric electric solution with the de Sitter
center.

\end{abstract}


]

\vskip0.1in

{\bf Introduction -} Seventy years ago Born and Infeld proposed a
Nonlinear Electrodynamics  starting from  {\it the principle of
finiteness: a satisfactory theory should avoid letting physical
quantities become infinite}\cite{born}. In the standard picture
based on the conception of a point charge, infinities come from
the fact that its self-energy is infinite. Born and Infeld
obtained a finite total energy for the field around a point charge
by introducing an upper limit on the electric field strength
$E_{max}$ connected with an electron finite radius by
$r_0=\sqrt{e/E_{max}}$ which gives the boundary of applicability
of their theory \cite{born}.

Today  NED theories appear as low energy effective limits in
certain models of string/M-theory (for review
\cite{witten,tseytlin}).

 In NED coupled to General Relativity there exists several
regular solutions describing  electrically charged black holes
\cite{garcia1,garcia2,garcia3,breton,garcia4}, magnetic black holes and monopoles
\cite{kirill}, and regular black holes of hybrid type which are electrically charged
but contain a magnetically charged core \cite{sasha}. There exist also a number of
theorems of non-existence (\cite{kirill} and references therein) prohibiting existence
of regular NED structures with an electric charge. They state that in NED with any
${\cal L}(F)$ such that ${\cal L}\sim{F}$ at small $F$ (the Maxwell weak field limit)
static spherically symmetric configurations with a non-zero electric charge cannot
have a regular center \cite{kirill}.

Another doubt about regular electrically charged structures
concerns the existence of singularities in the effective geometry
in which NED photons propagate \cite{mario}.

\vskip0.1in

The aim of this paper is to show that Nonlinear Electrodynamics
coupled to General Relativity and satisfying  the Weak Energy
Condition  guarantees existence of electrically charged regular
structures and provides a  cutoff on self-energy which diverges
for a point charge. One should only discard requirement of Maxwell
weak field limit at the center, on which non-existence theorems
are based, because a field must not be weak to be regular.

We prove the existence of electrically charged structures with the regular center, in
which geometry, field, and stress-energy tensor are regular without Maxwell limit as
$r\rightarrow 0$. We study propagation of photons in the effective geometry and
present the particular example of the new regular electrically charged solution.

The presented results apply directly to the cases when relevant NED scale is much
lower than the Planck scale.

In NED coupled to GR each electric solution has its magnetic
counterpart \cite{kirill}, whose existence is not forbidden, and
we concentrate here on the electric structure with its existential
problems.

\vskip0.1in

{\bf Energy Conditions -} The Weak Energy Condition (WEC),
$T_{\mu\nu}\xi^{\mu}\xi^{\nu}\geq 0$ for any timelike vector
$\xi^{\mu}$, which is satisfied if and only if \cite{HE}
$$
\rho\geq 0;  ~~ ~\rho + p_k \geq 0, ~~ k=1,2,3
                                                               \eqno(1a)
$$
guarantees that the energy density as measured by any local
observer is non-negative.

The Dominant Energy Condition  (DEC), $T^{00}\geq|T^{ik}|$ for
each $i,k=1,2,3$, which holds if and only if \cite{HE}
$$
\rho\geq0;~~\rho + p_k \geq 0;~~\rho-p_k \geq 0;~~~k=1,2,3
                                                              \eqno(1b)
$$
includes WEC and requires each principal pressure $p_k=-T^k_k$
never exceed the energy density which guarantees that speed of
sound cannot exceed the speed of light.

The Strong Energy Condition (SEC)  requires \cite{HE}
$$
\rho + \sum{p_k} \geq 0
                                                             \eqno(1c)
$$
and defines the sign of the gravitational acceleration.

\vskip0.2in

{\bf Symmetry of a source term -}
 Spherically symmetric electromagnetic field  with an arbitrary
gauge-invariant Lagrangian ${\cal L}(F), F=F_{\mu\nu}F^{\mu\nu}$,
has stress-energy tensor with the algebraic structure
$$
T^t_t=T^r_r
                                                                    \eqno(2)
$$
It is invariant under rotations in the $(r,t)$ plane, which enables to identify it as
a vacuum defined by the symmetry of its stress-energy tensor \cite{me92,me2000}. An
observer moving through a medium with stress-energy tensor of structure (2),  cannot
measure his velocity with respect to it which is typical for motion in a vacuum
\cite{gliner,landau}.

For the class of regular spherically symmetric geometries with the symmetry of a
source term given by (2), the Weak Energy Condition leads inevitably to de Sitter
asymptotic at approaching a regular center \cite{me2002}.

The basic fact of any geometry with de Sitter center generated by a source term of
type (2), which does not depend on whether we identify it as a vacuum or not, is that
the ADM mass of an object is related to both de Sitter vacuum trapped inside and
smooth breaking of space-time symmetry from the de Sitter group in the origin to the
Poincar\'e group at infinity \cite{me2002}.

For the spherically symmetric stress-energy tensor with the algebraic structure (2)
the equation of state relating density $\rho=T^t_t$ with the radial pressure
$p_r=-T^r_r$ and tangential pressure
$p_{\perp}=-T_{\theta}^{\theta}=-T_{\phi}^{\phi}$, reads \cite{me92,me2002}
$$
p_r=-\rho; ~~~ ~p_{\perp}=-\rho - \frac{r}{2}\rho^{\prime}
                                                                 \eqno(3)
$$
where prime denotes differentiation with respect to $r$.

\vskip0.2in

{\bf Basic equations -}
 In nonlinear electrodynamics minimally
coupled to gravity, the action is given by (in geometrical units
$G=c=1$)
$$
S=\frac{1}{16\pi}\int{d^4 x\sqrt{-g}({R-{\cal L}(F)}}); ~~ ~
F=F_{\mu\nu}F^{\mu\nu}
                                                                    \eqno(4)
$$
Here $R$ is the scalar curvature, and
$F_{\mu\nu}=\partial_{\mu}A_{\nu}-\partial_{\nu}A_{\mu}$ is the electromagnetic field.
The gauge-invariant electromagnetic Lagrangian ${\cal L}(F)$ is an arbitrary function
of $F$ which should have the Maxwell limit, ${\cal L} \rightarrow F,~ {\cal
L}_F\rightarrow 1$ in the weak field regime.

The action (2) gives the dynamic field equations
$$
({\cal L}_F F^{\mu\nu})_{;\mu}=0;  ~~~ ~ ^{*}F^{\mu\nu}_{ ~~;\mu}=0
                                                                   \eqno(5)
$$
where ${\cal L}_F=d {\cal L}/dF$. In the spherically symmetric
case the only essential components of $F_{\mu\nu}$ are a radial
electric field $F_{01}=-F_{10}=E(r)$ and a radial magnetic field
$F_{23}=-F_{32}$.

The Einstein equations take the form \cite{kirill}
$$
G^{\mu}_{\nu}=-T^{\mu}_{\nu}=2 {\cal L}_F
F_{\nu\alpha}F^{\mu\alpha}-\frac{1}{2}\delta_{\nu}^{\mu} {\cal L}
                                                                      \eqno(6)
$$
Definition of $T_{\mu\nu}$ here differs from standard definition
(see, e.g., \cite{landau}) by $8 \pi$, so that
$T^0_{0^(here)}=8\pi\rho$, etc.

The density and pressures ~for electrically charged structures are given by
$$
\rho=-p_r=\frac{1}{2}{\cal L}-F{\cal L}_F;~~~
p_{\perp}=-\frac{1}{2}{\cal L}
                                                                \eqno(7)
$$
and scalar curvature is
$$
R=2({\cal L}-F{\cal L}_F)=2(\rho - p_{\perp})
                                                               \eqno(8)
$$
 Symmetry of a source term (2)
leads to the metric
$$
 ds^2=g(r) dt^2 - \frac{dr^2}{g(r)} - r^2 d\Omega^2
                                                                \eqno(9)
 $$
where $d\Omega^2$ is the line element on a unit sphere. The metric
function and mass function are given by
$$
g(r)=1-\frac{2{\cal M}(r)}{r}; ~~~
 {\cal
M}(r)=\frac{1}{2}\int_0^r{\rho(x)x^2dx}
                                                                \eqno(10)
$$

Dynamical equations (5) yield
$$
r^2{\cal L}_F F^{01}=q,
                                                               \eqno(11)
$$
where $q$ is constant of integration identified as an electric
charge by asymptotic behavior in the weak field limit.

As follows from (10),
$$
F=2F_{01}F^{01}=-\frac{2q^2}{{\cal L}_F^2 r^4}
                                                            \eqno(12)
$$

Theorems of non-existence require the Maxwell behavior at the
regular center, ${\cal L}\rightarrow 0, {\cal L}_F\rightarrow 1$
as $F\rightarrow 0$. The proof is that regularity of stress-energy
tensor requires $|F{\cal L}_F| < \infty$ as $r\rightarrow 0$,
while $F{\cal L}_F^2 \rightarrow -\infty$ by virtue of (12), it
follows that ${\cal L}_F\rightarrow \infty$ and $F\rightarrow 0$,
which is strongly non-Maxwell behavior \cite{kirill}.

This sentence reads that a regular electrically charged structure
does not compatible with the Maxwell weak field limit ${\cal
L}\rightarrow 0, {\cal L}_F\rightarrow 1$ as $F\rightarrow 0$, in
the center.

However, if the density does not vanish as $r\rightarrow 0$, then
${\cal L}$ must not vanish there, although $F$  vanishes in all
cases of the regular center. Moreover, for solutions satisfying
the Weak Energy Condition, density takes maximum there, since the
WEC requires, by (1a) and (3), $\rho^{\prime}\leq 0$. Then $\rho$
is maximal at the center, and one cannot  expect validity of the
weak field limit in the region of maximal energy density of the
field.

Let us fix the basic properties of electrically charged NED configurations obligatory
for any Lagrangian ${\cal L}(F)$:

i) First is the fundamental observation of Bronnikov's theorem
\cite{kirill} - that $F$ must vanish as $r\rightarrow 0$ to
guarantee regularity, and  the electric field strength is zero in
the center of any regular electrically charged NED structure.

ii) Second is another observation from \cite{kirill} - since $F$ vanishes at both zero
and infinity where it should follow the Maxwell weak field limit, $F$ must have at
least one minimum in between where an electric field strength has a maximum. This
leads to branching of ${\cal L}(F)$ as a function of $F$. This inevitable feature  of
electrically charged solutions creates problems in an effective geometry whose
geodesics are world lines of NED photons \cite{mario}.

iii) Third is the existence of surface of zero gravity at which
Strong Energy Condition  is violated. For all electric NED
configurations this reads $2p_{\perp}=-{\cal L} \geq 0$, and SEC
is violated at the surface ${\cal L}=0$.

\vskip0.2in

{\bf NED structures satisfying WEC -} The Weak Energy Condition
requires density be non-zero and maximal in the origin, since with
$\rho \geq 0$ and $\rho^{\prime}\leq 0$, a density  cannot
decrease beyond zero being obliged to be non-negative. Combined
with the first property this raises the question - whose energy
density is maximal in the center of structures  where electric
field tension vanishes?

The basic feature of all solutions of class (2)  is de Sitter
behavior at approaching the regular center \cite{me2002}. Indeed,
 regularity of $\rho(r)$ requires $r\rho^{\prime}/2 \rightarrow 0
$ as $r\rightarrow 0$ (which is easily to check by taking $r\rho^{\prime}=$const and
calculating $\rho$). With $|\rho^{\prime}| < \infty$ the equation of state, by (3),
tends to $p_r=p_{\perp}=-\rho$  as $r\rightarrow 0$ which gives de Sitter asymptotic
$$
g(r)=1-\frac{\Lambda}{3} r^2
                                                              \eqno(13)
$$
with cosmological constant $\Lambda = 8\pi \rho(0)$. For electric
NED structure Lagrangian ${\cal L}(F)\rightarrow 2\rho(0)$ as
$r\rightarrow 0$, by (7), so that Lagrangian is positive and takes
its maximal value at the center which testifies that the limiting
density as $r\rightarrow 0$ is of electromagnetic origin.

Here we can answer the question {\it whose density is maximal as $r\rightarrow 0$
where electric field  vanishes.} The $T_0^0$ component of electromagnetic
stress-energy tensor  does not vanish (neither diverges) as $r\rightarrow 0$ and
provides an effective cutoff on self-interaction by relating it, through Einstein
equations, with cosmological constant $\Lambda$ corresponding to energy density of a
vacuum, in this case the electromagnetic vacuum (2).

The WEC requirement $\rho+p_{\perp} \geq 0$ leads to $-F{\cal L}_F
\geq 0$. It gives ${\cal L}_F \geq 0$. It gives also a constraint
on a Lagrangian $ ~{\cal L} \geq 2 F{\cal L}_F~$ as its obligatory
low boundary.

The DEC requirement $\rho-p_{\perp} \geq 0$, is satisfied when
${\cal L} \geq F{\cal L}_F$. With $F{\cal L}_F \leq 0$ by WEC,
this constraint is satisfied in the whole region surrounding the
center including a certain region outside the surface of violation
of the strong energy condition at which ${\cal L}=0$.

\vskip0.1in

Electrically charged solutions are typically found in the alternative form of NED
obtained by the Legendre transformation: one introduces the tensor $P_{\mu\nu}={\cal
L}_F F_{\mu\nu}$ with its invariant $P=P_{\mu\nu}P^{\mu\nu}$ and consider
Hamiltonian-like function ${\cal H}(P)=2F{\cal L}_F-{\cal L}$ as a function of $P$;
the theory is then reformulated in terms of $P$ and  specified by ${\cal H}(P)$
\cite{plebanski}. P frame is related with F frame by \cite{plebanski}
$$
{\cal L}=2P{\cal H}_P-{\cal H};  ~~~ {\cal L}_F{\cal
H}_P=1;~~F=P{\cal H}_{P}^2
                                                        \eqno(14)
$$
Here ${\cal H}_P=d{\cal H}/dP$. The electric invariant is
$$
P=-2P_{01}P^{01}=-\frac{2q^2}{r^4}
                                                          \eqno(15)
$$
The metric in P frame is calculated from (10) with
$$
\rho(r)=-\frac{1}{2}{\cal H}
                                                           \eqno(16)
$$

FP duality coincides with conventional electric-magnetic duality only in the Maxwell
limit where ${\cal L}=F=P={\cal H}$ \cite{kirill}. Interpretation  of the results
obtained in P framework depends essentially on transformation to F framework where
Lagrangian dynamics is specified. The two frames are equivalent only when the function
$F(P)$ is monotonic \cite{kirill}.

The function $F(P)$ which vanishes at both center and infinity has
at least one minimum in which
$$
{\cal L}_{FF}=\frac{1}{2}\biggl(\frac{{\cal H}_P}{F_P}-{\cal
L}_F\biggr)
$$
tends to infinities of opposite signs and ${\cal L}(F)$ suffers
branching. Additional branching is related to extrema of the
function ${\cal H}(P)$ \cite{kirill}.

While the first kind of branching is inevitable, the second is avoided by WEC, since
${\cal L}_F \geq 0$ results in ${\cal H}_P \geq 0$. When ${\cal H}(P)$ is monotonic
function, the function ${\cal L}(F)$ has only two branches  related to one minimum of
$F$ \cite{kirill}. This looses problems with restoring F-frame Lagrangian dynamics
from P-frame results. With one cusp interpretation is transparent and inevitable cusp
becomes the source of information about most interesting behavior of electrically
charged NED structures which displays in propagation of photons in an effective
geometry. Typical behavior of Lagrangian $\cal L$ as a function of $F$ is depicted in
Fig.1.
\begin{figure}
\vspace{-8.0mm}
\begin{center}
\epsfig{file=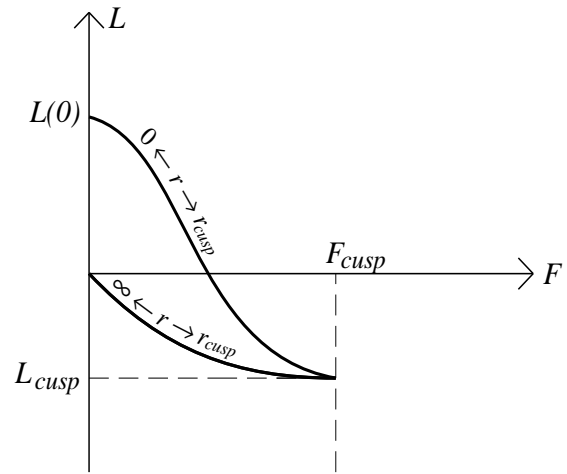,width=8.0cm,height=7.0cm}
\end{center}
\caption{ Typical behavior of a Lagrangian ${\cal L}(F)$.} \label{fig.1}
\end{figure}

 At the cusp surface $r=r_{cusp}$ the invariant $F$ has
  minimum as a function of $P$ and a function of $r$
 (since $P(r)$ is monotonic).
 The Lagrangian $r$-derivative there ${\cal L}^{\prime}={\cal L}_F
F^{\prime}=0$ and ${\cal L}$ takes its minimal value. The Lagrangian ${\cal L}(F)$
which is monotonic function of $F$ (${\cal L}_F \geq 0$), first decreases smoothly
along the first branch from its maximal value ${\cal L}(0)$ to ${\cal L}_{cusp}$ as
$F$ decreases from $F=-0$ at $r=0$ to $F_{min}=F_{cusp}$; then the Lagrangian
increases along the second branch from its minimal value ${\cal L}_{cusp} < 0$ to its
Maxwell limit ${\cal L}\rightarrow F\rightarrow -0$ as $F$ increases from  $F_{cusp}$
to $F\rightarrow -0$ as $r\rightarrow \infty$.

 At the cusp the electric field $E^2(r)=-F(r)/2$ achieves its
maximum.

Tangential pressure  is maximal at the cusp surface, where
$p_{\perp}^{\prime}=-{\cal L}^{\prime}/2=0$. In one-cusp
configurations tangential pressure has one extremum, this is
actually dictated by WEC which defines also the number of
horizons.  The function $g(r)$ has  only one minimum and geometry
described by the  metric (9) can have not more than two horizons
\cite{me2002}.

With ${\cal H}_P\geq 0$ the electric susceptibility $\epsilon=1/{\cal H}_P$ is
everywhere positive. When $P\rightarrow -\infty$ at the center ${\cal H}_P\rightarrow
+0$ (since ${\cal L}_F \rightarrow \infty$ there),  this leads to  $\epsilon
\rightarrow +\infty$, electric susceptibility is divergent, so that electrically
charged NED configurations demonstrate ideal conducting behavior at approaching the
regular center where the electric field tension vanishes.

\vskip0.1in

Summarizing we conclude that {\it regular electrically charged NED
structures satisfying Weak Energy Condition,  have de Sitter
center, not more than two horizons and precisely one cusp of
${\cal L}(F)$ where the electric field strength achieves its only
maximum.}

\vskip0.2in

{\bf New exact electric NED solution -} Let us choose the function
${\cal H}(P)$ in the form
$$
{\cal H}(P)=\frac{P}{(1+\alpha \sqrt{-P})^2}
                                                            \eqno(17)
$$
where $\alpha$ is characteristic parameter of the NED theory. Then
we get
$$
{\cal H}_P=\frac{1}{(1+\alpha \sqrt{-P})^3}
                                                        \eqno(18)
$$
$$
F=\frac{P}{(1+\alpha \sqrt{-P})^6}; ~~~~
F_P=\frac{(1-2\alpha
\sqrt{-P})}{(1+\alpha \sqrt{-P})^7}
                                                          \eqno(19)
$$
  With $P$ defined by (15) this gives
  $$
  {\cal H}=-\frac{2q^2}{(r^2+r_0^2)^2};~ ~~~
{\cal H}_P= \frac{r^6}{(r^2+r_0^2)^3} ~
                                                      \eqno(20)
$$

The parameter $r_0^2=\alpha \sqrt{2q^2}$ is fixed by integrating
(10) with the density (16) which connects $r_0$ with the total
mass $m={\cal M}(r\rightarrow \infty)$. This gives
$$
r_0=\frac{\pi}{8}\frac{q^2}{m}
                                                                \eqno(21)
$$
as classical electromagnetic radius modified by numerical coefficient of chosen
particular NED model (17).

 The only minimum of
$F(P)$ is at $2\alpha \sqrt{-P}=1$ and the cusp surface is given
by
$$
r_{cusp}=\sqrt{2} r_0
                                                                 \eqno(22)
$$
The density and pressure are (up to $8\pi$ mentioned above)
$$
  \rho(r)=\frac{q^2}{(r^2+r_0^2)^2};~~~~
 p_{\perp}=\frac{q^2 (r^2-r_0^2)}{(r^2+r_0^2)^3}
                                                                 \eqno(23)
  $$
Function $\rho(r)$ is monotonically decreasing, function
$p_{\perp}(r)$ achieves maximum at the cusp surface.

The electric field  is given by
$$
 F=-\frac{2q^2 r^8}{(r^2+r_0^2)^6};~~~
 E^2=\frac{q^2 r^8}{(r^2+r_0^2)^6}
                                                              \eqno(24)
$$
It achieves its maximum at the cusp surface
$$
E_{max}=\frac{4}{27}\frac{q}{r_0^2}
                                                             \eqno(24a)
$$
and  Maxwell limit $E\rightarrow 0$ as $r\rightarrow \infty$.

Lagrangian and its derivative are
$$
{\cal L}=\frac{2q^2(r_0^2-r^2)}{(r^2+r_0^2)^3}; ~~~ {\cal
L}_F=\frac{(r^2+r_0^2)^3}{r^6}
                                                            \eqno(25)
$$

The scalar curvature for this Lagrangian
 is given by
$$
R=\frac{4q^2 r_0^2}{(r^2+r_0^2)^3}
                                                              \eqno(26)
$$
It is positive everywhere, and the Dominant Energy Condition is
satisfied (although we did not impose it) which is the good
feature, since e.g., propagation of NED photons in an effective
geometry resembles propagation inside a dielectric medium
\cite{mario}, and DEC makes it free of effects produced by speed
of sound exceeding speed of light.

Integrating (10) with the density profile (23) we get the metric
$$
g(r)=1-\frac{4m}{\pi r}\biggl(arctg \frac{r}{r_0}-\frac{r
r_0}{r^2+r_0^2}\biggr)
                                                                 \eqno(27)
$$
For $r \gg r_0$ it reduces to
$$
g(r)=1-\frac{2m}{r}+\frac{q^2}{r^2}-\frac{2}{3}\frac{q^2
r_0^2}{r^4}
                                                              \eqno(28)
$$
and has Reissner-Nordstr\"om limit as $r\rightarrow \infty$.

At small values of $r$, $r \ll r_0$ we get de Sitter asymptotic
(13)  with the cosmological constant
$$
\Lambda=\frac{q^2}{r_0^4}
                                                              \eqno(29)
$$
which gives proper expression for a cutoff on self-energy density
by the finite value of electromagnetic density $T_0^0
(r\rightarrow 0)$ related to the cosmological constant $\Lambda
=8\pi \rho(0)$ which appears at the regular center.

The mass, of electromagnetic origin, is related to this cutoff by
$m=\pi^2 \rho(0) r_0^3$, where $r_0$ is the classical
electromagnetic radius.

\vskip0.1in

Characteristic parameter which decides if a solution describes a
regular electrically charged black hole either self-gravitating
particle-like structure with de Sitter vacuum inside, is given by
 $$
 \beta=\frac{8}{\pi^2}\biggl(\frac{2m}{q}\biggr)^2=\frac{2}{\pi}\frac{r_g}{r_0}
                                                        \eqno(30)
 $$
where $r_g=2m$ is the characteristic Schwarzschild radius.

For $\beta > \beta_{crit}=2.816$ solution describes a black hole.

 For $r_g\gg r_0$, two horizons are
$$
r_{-}\simeq {r_{S}\biggl(1+1.4 \frac{r_0}{r_g}\biggr)};~~~
r_{+}\simeq{r_g\biggl(1+1.3
\frac{r_0}{r_g}\biggr)}
                                                           \eqno(31a)
$$
Internal horizon in this limit is close to de Sitter horizon $r_S=\sqrt{3/\Lambda}$,
and an event horizon to the Schwarzschild horizon $r_g$.

 For $\beta =\beta_{crit}$ there is a double horizon
 $$
 r_{\pm}=1.825~ r_0
                                                             \eqno(31b)
$$
The global structure of  space-time with horizons is precisely the
same as for de Sitter-Schwarzschild geometry \cite{me96} (pictures
for two and one horizon cases are presented in \cite{us2003},
Fig.3), and differs from Reissner-Nordstr\"om case only in that
the timelike surface $r=0$ is regular.

In terms of $q/2m$ black hole exists for $q/2m \leq 0.536$, and
for $q/2m > 0.536$ we have electrically charged self-gravitating
particle-like NED structure.

\vskip0.2in

{\bf NED electrically charged structure from the point of view of
photons -} In nonlinear electrodynamics photons do not follow null
geodesics of background geometry, but propagate along null
geodesics of an effective geometry \cite{effective}. In the
spherically symmetric case it is described by
 the metric \cite{mario}
$$
ds^2_{eff}=\frac{g(r)}{\Phi(r)}dt^2-\frac{dr^2}{g(r)\Phi(r)}-\frac{r^2}{{\cal
L}_F(r)}d\Omega^2
                                                                  \eqno(32)
$$
where
$$
\Phi(r)=\frac{{\cal H}_P}{F_P}={\cal L}_F+2{\cal L}_{FF}
                                                                 \eqno(33)
$$

For any NED background geometry satisfying WEC, ${\cal H}_P\geq 0$
everywhere, and $F$ has one minimum. The function $\Phi$ is
negative for $r < r_{cusp}$ where $F$ decreases, and positive for
$r > r_{cusp}$ where $F$ increases, so that $\Phi \rightarrow
+\infty$ when $r\rightarrow r_{cusp}+0$\footnote{For the
 above exact solution $\Phi(r)=\frac{(r^2+r_0^2)^4}{r^6(r^2-2r_0^2)}$.}.
 At infinity, where ${\cal H}=P=F={\cal L}$, we get $\Phi
\rightarrow 1$ and ${\cal L}_F \rightarrow 1$. So, for a distant
observer an effective metric is close to his background metric.

One cusp in background geometry  creates one problem in the
effective geometry. The most essential consequence of a cusp is
the redshift as measured by a distant observer which is given by
\cite{mario}
$$
1+z=\frac{\Phi}{\sqrt{g}}
                                                           \eqno(34)
$$
It diverges at the BH horizon where $g(r)$ vanishes, and at the
cusp surface $r=r_{cusp}$ where $\Phi$ diverges.

 For a distant observer photons disappear beyond the surface
 $r=r_{cusp}$ in the same way as they disappear beyond the event
 horizon of a black hole.

 To investigate what is going with photons after crossing the cusp
 surface, we should change coordinates $r,t$ (which exchange their
 roles after crossing $r_{cusp}$), to coordinates $R, \tau$ given
 by
$$
R=t+\int{\frac{dr}{g(r)\sqrt{1-g(r)/\Phi(r)}}}
                                                             \eqno(35a)
$$
$$
\tau =t +
\int{\biggl(\frac{1}{g}-\frac{1}{\Phi}\biggr)\frac{dr}{\sqrt{1-g/\Phi}}}
                                                            \eqno(35b)
$$

In the case of background geometry without horizons the metric
function $g(r)$ is everywhere positive. We concentrate now on this
case to show how photons feel a cusp itself, when there is no
other peculiarities in space-time which they penetrate. (Their
behavior on horizons related to $g(r)$ is similar to the case
without a cusp.)

In coordinates $(R, \tau)$ the metric (32) transforms to
$$
ds^2_{eff}=d\tau^2-\biggl(1-\frac{g}{\Phi}\biggr)dR^2
-\frac{r^2(R, \tau)}{{\cal L}_F(r(R, \tau))}d\Omega^2
                                                            \eqno(36)
$$
The  function $g_{RR}$ never change its sign. The function $\Phi$
achieves its minimal positive value $\Phi \rightarrow 1$ as
$r\rightarrow \infty$ (where (32) coincides with the background
metric).

 In coordinates  $(R, \tau)$ surfaces $r=const$ are
straight lines $\tau - R =const$ (more precise,  some $f(r=const)$, see, e.g.,
\cite{landau}). Equation of light cones are given by
$$
\frac{d\tau}{dR}=\pm\sqrt{1-\frac{g}{\Phi}}
                                                        \eqno(37)
$$

 At infinity $g\rightarrow 1, \Phi\rightarrow 1$ and $d\tau/dR \rightarrow 0$,
 the cones are entirely open (compare
with the Schwarzschild case where $d\tau/dR =\pm \sqrt{r_g/r}$). Light cones in an
effective geometry (36) are shown in Fig.2.
\begin{figure}
\vspace{-8.0mm}
\begin{center}
\epsfig{file=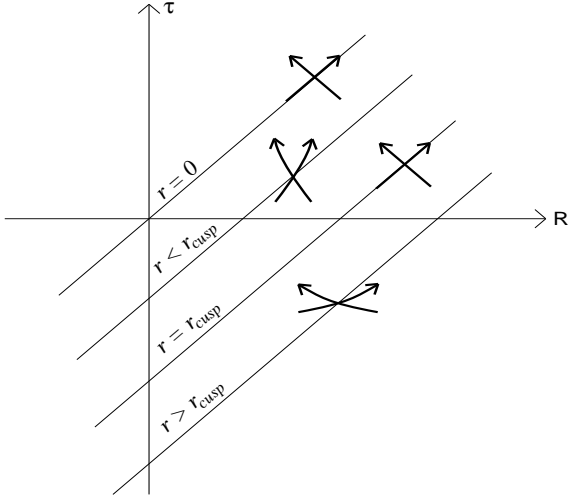,width=8.0cm,height=7.0cm}
\end{center}
\caption{Light cones (37) in an effective geometry (36).} \label{fig.1}
\end{figure}
For $r > r_{cusp}, \frac{d\tau}{dR} < 1$, outgoing photons move towards a distant
observer, ingoing photons towards a center.

For $r=r_{cusp}, \frac{d\tau}{dR}=\pm 1$, ingoing photons moves
toward a center, outgoing photons are kept on the surface
$r=r_{cusp}$ (as it should be for a horizon which is the null
surface, in Schwarzschild case $\frac{d\tau}{dR}=\pm
\sqrt{\frac{r_g}{r}}=\pm 1$).

For $r < r_{cusp}$ cones are directed inward - all photons are
ingoing. The cone is maximally narrow at the surface where
$1-\frac{g}{\Phi}$ is maximal, this is near the surface of zero
gravity (for considered solution, $r\simeq r_0$). Beyond this
surface cones start to widen again, still being inward-oriented
(actually this is $T_{-}$ region, a black hole in the effective
geometry).

At approaching the center, ${\cal H}_P={\cal L}_F^{-1} \rightarrow
0$ (since ${\cal L}_F \rightarrow \infty$ as $r\rightarrow 0$ for
any electrically charged solution), while $F_P < \infty$.
Therefore $\frac{d\tau}{dR}\rightarrow \pm 1$ as $r\rightarrow 0$.
Some of photons are kept on the surface $r=0$! Let us compare with
Schwarzschild case where $\frac{d\tau}{dR} =\pm
\sqrt{\frac{r_g}{r}}\rightarrow \infty$ as $r\rightarrow 0$, cones
are entirely closed and perpendicular to the surface $r=0$ which
is the space-like singularity of the Schwarzschild geometry. In
space-time created by NED electrically charged structure for
photons, one of ruling of a light cone is on the surface $r=0$,
which appears to be the {\it null surface} of the effective
geometry.

 The further analysis is straightforward -
one changes signs in (35b) to map $T_{+}$ region (a white hole in the effective
geometry), and introduces Kruskal coordinates to follow photons everywhere and
construct global structure of space-time as seen by NED photons.

 The main conclusion is that photons world lines never
terminate which suggests the absence of singularities in the
background geometry satisfying WEC.

\vskip0.2in

{\bf NED and cosmological constant -} In Quantum Field Theory relativistic field is
considered as collection of harmonic oscillators of all possible frequencies, and
vacuum as a superposition of ground states of all fields with the energy $E_{vac}=\sum
\frac{1}{2}\hbar\omega$, where $\frac{1}{2}\hbar\omega$ is a zero-point energy of each
particular field mode. Zero-point vacuum contribution to a stress-energy tensor has
the form $ <T_{\mu\nu}>=<\rho_{vac}>g_{\mu\nu}$ and behaves like a cosmological term.
In QFT an upper cutoff on a vacuum energy density is estimated by a scale at which our
confidence in formalism of QFT is broken, $E_{Pl}\sim{10^{19}}$ GeV. With
$\rho_{vac}\sim{\rho_{Pl}}\sim{10^{93}} g~cm^{-3}$, a QFT zero-point contribution to
gravity through Einstein equations, is incompatible with observational data which
constraint the vacuum contribution by 70 \% of total density in the Universe,
$\rho_{total}\sim{10^{-30} g ~cm^{-3}}$.

This creates the {\it problem of cosmological constant} \cite{weinberg}  which traces
back directly to divergent self-energy of a point charge. In QED the infinite
electromagnetic mass connected with self-energy of a point charge, is renormalized to
a finite observable value of $m_e$ by introducing an equally infinite negative mass of
non- electromagnetic (and actually still unknown) origin \cite{okun}.

As it was noted already in \cite{born},  theories based on geometrical assumptions
about the "shape" of charge, etc, break down on similar reason: they are compelled to
introduce {\it cohesive forces} of non-electromagnetic origin.

From above analysis it follows that NED coupled to GR makes it possible to get a
regular extended charged structure without involving cohesive forces of
non-electromagnetic origin. Finite self-energy is of purely electromagnetic origin
since it comes to Einstein equations from electromagnetic Lagrangian, but then
gravity, through Einstein equations, transforms  {\it cohesive forces of
electromagnetic origin} (negative pressure), into repulsive gravitational forces
acting beyond the surface at which strong energy condition is violated. Beyond this
surface NED structure displays rather de Sitter than Maxwell behavior, with vacuum
energy density $\Lambda$ coming from GR equations as an effective cutoff on a
self-energy.

Summarizing we conclude that {\it NED coupled to GR and satisfying
WEC, provides at the classical level a cutoff on self-energy
related to cosmological constant $\Lambda$.}

\vskip0.1in

At the classical level a NED structure with de Sitter center can be considered as an
elementary excitation of an electromagnetic vacuum (2).  A simplest quantum version
can be obtained in frame of a minisuperspace model. A spherical object with de Sitter
vacuum trapped inside can be described by a minisuperspace model with a single degree
of freedom. Quantization leads to the Wheeler-De Witt equation which reduces to the
Schr\"odinger equation with the potential $V(r)=-{\cal M}(r)/r$ \cite{me2002}. Near
the minimum of the potential, Schr\"odinger equation in turn reduces to the equation
for the harmonic oscillator whose basic frequency gives the energy of a zero-point
vacuum mode $E_0\sim {\hbar \sqrt{\Lambda}}$ which never exceeds the absolute value of
the negative binding energy $V_{min}$. One can expect that a cutoff on a self-energy
existing in the classical model would survive in its quantum version.

\vskip0.2in

{\bf Conclusion -} In Nonlinear Electrodynamics coupled to General Relativity and
satisfying the Weak Energy Condition, regular electrically charged structures exist
and have de Sitter regular center. In de Sitter center of a NED structure, electric
field, geometry and stress-energy tensor are regular without Maxwell limit which is
replaced by de Sitter limit: while the electric field tension vanishes, energy density
of a field takes its maximal value (vacuum density in the center) which gives an
effective cut-off on self-energy density, produced by NED coupled to gravity and
related to cosmological constant $\Lambda$. Mass of objects which is of purely
electromagnetic origin, is related to both de Sitter vacuum trapped inside and smooth
breaking of space-time symmetry from the de Sitter group in the origin to the
Poincar\'e group at infinity. All this concerns also structures possessing magnetic
charge or both, since in NED coupled to GR each electric solution has its magnetic
counterpart.

\vskip0.2in

\centerline{\bf Acknowledgment}

This work has been supported at the final stage by the Polish Committee for Scientific
Research through the grant  1 P03D 023 27.

\end{document}